\documentclass[aps,prd,superscriptaddress,twocolumn]{revtex4}

\usepackage{amsmath,amssymb}
\usepackage{color}
\usepackage{hyperref}

\newcommand{\bea}{\begin{eqnarray}}
\newcommand{\eea}{\end{eqnarray}}
\newcommand{\be}{\begin{equation}}
\newcommand{\ee}{\end{equation}}
\newcommand{\nn}{\nonumber}

\def\a{\alpha}      
\def\b{\beta}       
\def\c{\gamma} \def\g{\gamma}    
\def\d{\delta}    
        
\def\f{\phi}

\def\m{\mu} \def\n{\nu}

\def\t{\tau}
\def\th{\theta}

\def\hs{\hspace}
\def\p{\partial}

\def \be {\begin{equation}}
\def \ee {\end{equation}}
\def \ba {\begin{array}}
\def \ea {\end{array}}
\def \bea{\begin{eqnarray}}
\def \eea{\end{eqnarray}}

\def \a {\alpha}
\def \b {\beta}
\def \g {\gamma}

\def \d {\delta}

\def \m {\mu}
\def \n {\nu}

\def \O {\Omega}
\def \th {\theta}

\def \t {\tau}

\def \p {\partial}
\def \f {\frac}

\def \nn {\nonumber}

\def \ma {\mathcal}

\def \lt {\left}
\def \rt {\right}

\def \sr {\sqrt}

\def \hs {\hspace}

\def \Im {{\textrm{Im}}}

\begin{document}

\title{Electromagnetic duality in dyonic RN/CFT correspondence}

\author{Bin Chen}
\email[Email: ]{bchen01@pku.edu.cn}
\affiliation
{Department of Physics and State Key Laboratory of Nuclear Physics and Technology, Peking University, Beijing 100871, P. R. China}
\affiliation{Center for High Energy Physics, Peking University, Beijing 100871, P. R. China}
\author{Jia-ju Zhang}
\email[Email: ]{jjzhang@pku.edu.cn}
\affiliation
{Department of Physics and State Key Laboratory of Nuclear Physics and Technology, Peking University, Beijing 100871, P. R. China}

\begin{abstract}

The area law of Bekenstein-Hawking entropy of the black hole suggests that the black hole should have a lower-dimensional holographic description. It has been found recently that such holographic pictures could be set up from the study of the thermodynamics of both outer and inner horizons for a large class of rotating and charged black holes. For a four-dimensional dyonic Reissner-Nordstr\"om black hole, its thermodynamics indicates that it has multiple holographic pictures not only the electric and magnetic ones corresponding to the conserved electric and magnetic charges carried by the black hole, but also the more generic ones generated by the $SL(2,Z)$ transformations. We show that this $SL(2,Z)$ group originates from the underlying electromagnetic duality symmetry in the Einstein-Maxwell theory. It turns out that the thermodynamics of the black hole not only encodes the information of holographic pictures; moreover it could reflect the symmetries of the underlying theory.


\end{abstract}

\maketitle

\section{Introduction}

One of the central issues in quantum gravity is to understand the thermodynamics in black hole physics \cite{Bardeen:1973gs}, in particular the Bekenstein-Hawking entropy of a black hole \cite{Bekenstein:1973ur,Hawking:1974sw}. The entropy turns out to be proportional to the area of the horizon of the black hole, rather than the volume inside the horizon. This unexpected area law of entropy posed a serious problem of how to count a black hole's entropy in a microscopic way. In the middle of the 1990s, the entropy for a kind of extremal black hole in string theory could be reproduced successfully for the first time by counting the degeneracy of brane configurations forming the black hole \cite{Strominger:1996sh}. More precisely, the microscopical entropy is encoded in a two-dimensional (2D) conformal field theory (CFT), whose target is the moduli space of the brane configurations. This kind of counting has been generalized to the near extremal and even nonextremal cases. One important lesson obtained from these studies is that a black hole could be holographically described by a 2D CFT.

In the past few years, the holographic 2D CFT description of usual four-dimensional (4D) rotating and charged black holes which could not be embedded into string theory in a simple way has been proposed and tested \cite{K1,K2}. In this so-called Kerr/CFT correspondence, people used completely different techniques, including the asymptotic symmetry group analysis for the near horizon geometry of extremal black holes \cite{K1}, and the hidden conformal symmetry in the low-frequency scattering of a scalar off the nonextreme black hole \cite{K2}, to investigate the holographic dual of the black hole. As these techniques have nothing to do with string theory, they have been widely and successfully applied to the study of many kinds of black holes \cite{Lu:2008jk,Hartman:2008pb,Krishnan:2010pv,Chen:2010ywa,Chen:2011wm}. However, for the generic nonextreme case, it is still not clear how to determine convincingly the information of the dual CFT, including the central charges and the temperatures.

Very recently, new light has been shed on the black hole/CFT correspondence from the point of view of thermodynamics \cite{Chen:2012mh,Chen:2012yd,Chen:2012ps}. Surprisingly, it was found that much of the information of the dual 2D CFT is encoded in the thermodynamics of the black hole itself. The key role is played by the thermodynamics of the inner horizon, which has been ignored for a long time. From the first laws of thermodynamics of both the outer and inner horizons, one may compose the first laws of thermodynamics of a right-moving sector and a left-moving sector. It turns out that these two sectors differ from the right-moving and left-moving sectors of dual CFT only by a scale factor. The scale factor could be determined from the first laws by considering the basic ``quantized" unit of perturbation. In such a way, much of the information of the dual CFT, including the temperatures and central charges, could be read out easily. This method is remarkably effective, though intriguing. It has not only reproduced successfully the holographic pictures in well-established black hole/CFT correspondence \cite{Chen:2012mh}, predicted the holographic picture of doubly rotating black ring \cite{Chen:2012mh,Chen:2012yd}, but it also helped to clarify the puzzles in Reissner-Nordstr\"om (RN)/CFT correspondence \cite{Chen:2012ps}.

In the application of the thermodynamics method to set up the holographic pictures of the black hole, one advantage is that it allows us to read elementary CFT duals in a simple way, with each dual corresponding to a conserved charge. Moreover, it allows us to generate more generic dual pictures straightforwardly. For example, for the 4D Kerr-Newman case, there ia an elementary $J$ picture based on the $U(1)$ rotation and an elementary $Q$ picture based on $U(1)$ gauge symmetry \cite{Hartman:2008pb,Chen:2010ywa}, from which the more generic pictures could be generated by acting $SL(2,Z)$ transformations on the two elementary pictures. In this case, the $SL(2,Z)$ group originates from the modular group of the torus in the uplifted five-dimensional (5D) metric \cite{Chen:2011wm}. However, in some other cases, the origin of the transformation group is not clear.

In this paper, we investigate the holographic pictures of the dyonic RN black hole in 4D Einstein-Maxwell theory, using the thermodynamics method. This black hole carries both electric and magnetic charges. Correspondingly, it has two elementary holographic pictures, which we call the electric and magnetic pictures. Using the first laws of thermodynamics, we may generate other holographic pictures by enacting an $SL(2,Z)$ transformation on the elementary pictures. Unlike 4D Kerr-Newman or 5D Myers-Perry black holes, this $SL(2,Z)$ cannot be explained geometrically. Here we show that it originates from the electromagnetic duality of the Einstein-Maxwell theory.

\section{Dyonic Reissner-Nordstr\"om Black Hole}

The dyonic RN black hole is a solution of the Einstein-Maxwell theory with the action
$I=I_{EH}+I_{EM}$. $I_{EH}$ is the Einstein-Hilbert action
\be
I_{EH}=\f{1}{16\pi G} \int d^4x\sr{-g}R,
\ee
and $I_{EM}$ is the action of the Maxwell theory in a curved background
\be \label{iem}
 I_{EM}=-\f{1}{16\pi} \int d^4x\sr{-g} \lt( F_{\m\n}F^{\m\n}
+\f{\th e^2}{2\pi} F_{\m\n}*F^{\m\n} \rt)
\ee
with $*$ being the Hodge duality. We set $c=\hbar=1$, $G=\ell_p^2$, and use Gauss convention. To fully capture the features of electromagnetic duality, we have included the $\th$ term in the action. We may combine the two real coupling constants $e,\th$  into a complex coupling parameter,
\be
\t=\f{\th}{2\pi}+\f{i}{e^2}.
\ee

The metric of the black hole is of the form
\be \label{e15}
ds^2_4=-N^2 dt^2+g_{rr}dr^2+r^2 (d\th^2+\sin^2\th d\phi^2),
\ee
with
\be N^2=\f{1}{g_{rr}}= 1-\f{2 G M}{r}+\f{G Q^2}{r^2}. \ee
Here $M$ is the mass of the black hole, and
\be
Q^2=Q_e^2+Q_m^2,
 \ee
with $Q_{e,m}$ being the electric and magnetic charges, respectively. As there is no globally well-defined gauge field for a dyon, we need to introduce two gauge fields to cover the whole spacetime manifold:
\be
A_{N,S}=-\f{Q_e}{r}dt+Q_m (\cos\th \mp 1) d\phi.
\ee
The upper minus sign applies to the north half-sphere, and the lower plus sign applies to the south half-sphere.

The horizons are located at $r_\pm=GM\pm\sr{G^2M^2-GQ^2}$. The entropy product of the outer and inner horizons is
\be
\f{S_+ S_-}{4\pi^2}=\f{Q^4}{4},
\ee
which is mass independent.

There is electromagnetic duality in the action (\ref{iem}), which could be recast as
\be \label{lem}
I_{EM}=-\f{1}{32\pi} \int d^4x \sr{-g} \Im \ma F_{\m\n} \ma G^{\m\n},
\ee
with $\ma F_{\m\n}=e(F_{\m\n}+i*F_{\m\n})$, $\ma G_{\m\n}=\t \ma F_{\m\n}$. Under the $SL(2,Z)$ transformation
\be
\lt( \ba{c} \ma G'_{\m\n} \\ \ma F'_{\m\n} \ea \rt)=
\lt(\ba{cc} \a & \b \\ \c & \d \ea \rt)
\lt( \ba{c} \ma G_{\m\n} \\ \ma F_{\m\n} \ea \rt),
\ee
with $\a,\b,\c,\d \in Z$ and $\a\d-\b\c=1$, the action is invariant. From $\ma G'_{\m\n}=\t' \ma F'_{\m\n}$ it is easy to see that $\t$ transforms as
\be
\t'=\f{\a\t+\b}{\c\t+\d},
\ee
which is the feature of the electromagnetic duality.
Correspondingly, the charges transform as
\be \label{e19}
\lt( \ba{c} e' Q_{e'}\\ \f{Q_{m'}}{e'}-e' Q_{e'}\f{\th'}{2\pi} \ea \rt)=
\lt(\ba{cc} \d & -\c \\ -\b & \a \ea \rt)
\lt( \ba{c} e Q_{e}\\ \f{Q_{m}}{e}-e Q_{e}\f{\th}{2\pi} \ea \rt).
\ee
Due to Witten's effect \cite{Witten:1979ey}, the electric and magnetic charges should be
\bea \label{qeqm}
&& Q_e=N_e e-N_m \f{e\th}{2\pi},  \nn\\
&& Q_m=\f{N_m}{e},
\eea
with $N_{e,m}$ being integers and transforming as
\be
\lt( \ba{c} N_{e'}\\N_{m'} \ea \rt)=
\lt(\ba{cc} \a & \b \\ \c & \d \ea \rt)
\lt( \ba{c} N_{e}\\ N_{m} \ea \rt).
\ee

It can be verified easily that the quantity
\be
Q^2=Q_e^2+Q_m^2=\f{1}{\Im \t}|N_e-\t N_m|^2
\ee
is invariant under the above $SL(2,Z)$ transformation. Therefore the dyonic black hole geometry (\ref{e15}) does not change under the transformation of the electromagnetic duality. In other words, the geometry is the solution of different theories, which have different coupling constants and gauge potentials but are related to each other by the electromagnetic duality. This fact suggests that the dyonic black hole could be studied from the points of view of different theories, leading to different dual pictures.

\section{Dyonic RN/CFT correspondence}

The dyonic RN black hole (\ref{e15}) has the first laws of thermodynamics at the outer and the inner horizons as follows:
\bea
&&d M=T_+ d S_+ + \O_+^e d N_e + \O_+^m d N_m  \nn\\
&&\phantom{d M}=-T_- d S_- + \O_-^e d N_e + \O_-^m d N_m.
\eea
Here $M$ is the mass of the black hole, and $T_\pm$, $S_\pm$ are the temperatures and the entropies of the outer and inner horizons. $N_{e,m}$ are two integer-valued charges related to $Q_{e,m}$ through (\ref{qeqm}), and $\O_\pm^{e,m}$ are their corresponding intensive quantities at the outer and the inner horizons. As clarified in  \cite{Chen:2012ps}, the quantization condition is essential to set up the holographic dual without ambiguity such that the integer-valued quantities $N_e$ and $N_m$ have been used in the above relations. It could be proved that the first law of the outer horizon always indicates that of inner horizon under reasonable assumptions \cite{Chen:2012mh}, and  if $T_+S_+=T_-S_-$ is satisfied, then the entropy product $S_+S_-$ is mass independent.
For convenience, one can define an entropy product function
\be \label{e7}
\ma F\equiv \f{S_+ S_-}{4\pi^2},
\ee
with $\ma F$ being a function of the charges $N_{e,m}$ in our case.

After making the combinations \cite{Cvetic:1997uw,Cvetic:1997xv,Cvetic:2009jn}
\bea
&& T_{R,L}=\f{T_-T_+}{T_- \pm T_+},  \nn\\
&& S_{R,L}=\f{1}{2}(S_+ \mp S_-),  \label{SRL}\\
&& \O_{R}^{e,m}=\f{T_- \O_+^{e,m} + T_+ \O_-^{e,m}}{2(T_- + T_+)},  \nn\\
&& \O_{L}^{e,m}=\f{T_- \O_+^{e,m} - T_+ \O_-^{e,m}}{2(T_- - T_+)}, \nn
\eea
we can rewrite the first laws of the outer and inner horizons as those for the right- and  left-moving sectors:
\bea \label{e11}
&&\f{1}{2}d M=T_R d S_R+\O_R^e d N_e  +\O_R^m d N_m \nn\\
&&\phantom{\f{1}{2}d M}=T_L d S_L  +\O_L^e d N_e  +\O_L^m d N_m.
\eea
Keeping $N_m$ invariant, i.e., considering the perturbation of the type $(dN_e,dN_m)=dN_e(1,0)$  from the above first laws, we could get
\be \label{e6}
d N_e=\f{T_L}{\O_R^e-\O_L^e}d S_L  - \f{T_R}{\O_R^e-\O_L^e}d S_R.
\ee
Now we can read the information of the dual CFT. There are two remarkable facts concerning the above manipulations:
\begin{enumerate}
\item $S_{R,L}$ in (\ref{SRL}) are exactly the entropies of the right- and left-moving sectors of CFT dual to the black hole \cite{Cvetic:1996kv,Larsen:1997ge,Cvetic:1997uw,Cvetic:1997xv,Cvetic:2009jn}.
\item Keeping $N_m$ invariant, we have the $N_e$ picture; i.e., the electric picture.  There is the first law of thermodynamics
\be \label{e8}
d N_e=T_L^e d S_L-T_R^e d S_R,
\ee
with $T_{R,L}^e$ being exactly the right- and left-moving temperatures in the dual CFT \cite{Chen:2012mh,Chen:2012ps}.
\end{enumerate}
We assume that the CFT entropies could be reproduced by Cardy's formula
\be \label{cardy}
S_{R,L}=\f{\pi^2}{3}c_{R,L}^e T_{R,L}^e,
\ee
from which we may derive the central charges. It could be shown easily  that if $\cal F$ is mass independent then the right- and left-moving sector central charges must be equal; $c_R^e=c_L^e$ \cite{Chen:2012mh}. We substitute $S_\pm=S_L \pm S_R$ into (\ref{e7}) and take variations on both sides of the equation while keeping $N_m$ invariant, and we get
\be
\f{\p \ma F}{\p N_e}d N_e=\f{S_L d S_L-S_R d S_R}{2\pi^2}.
\ee
From the above equation, using the first law (\ref{e8}), Cardy's formula (\ref{cardy}), and the fact that $c_R^e=c_L^e$, we find
\be \label{ce}
c_{R,L}^e=6\f{\p \ma F}{\p N_e}=6eQ_e Q^2.
\ee
These are the central charges of dual CFT in the  electric picture \cite{Chen:2010yu}. Similarly, we could keep $N_e$ invariant and obtain the $N_m$ picture; i.e., the magnetic picture, in which the dual CFT is of that the central charges
\be \label{cm}
c_{R,L}^m=6\f{\p \ma F}{\p N_m}=6Q^2 \lt( \f{Q_m}{e}-e Q_e \f{\th}{2\pi} \rt).
\ee

The relation between the entropy product function $\ma F$ and the central charges is remarkable. For more general cases, there is always the relation \cite{Chen:2013rb}
\be
c^i_{R,L}=6\f{\p \ma F}{\p N_i}.
\ee
Here $N_i$ is one of the integer-valued charges appearing in the first laws, and could be angular momentum, or another conserved $U(1)$ charge.

There is a simple way to understand the above elementary pictures. Actually, the relation in (\ref{e11}) tells us how the black hole responds to various kinds of perturbations. If we consider the perturbations carrying only electric charges, or more precisely, $(d N_e,d N_m)=dN_e(1,0)$, the corresponding first law (\ref{e8}) gives us the temperatures $T^e_{R,L}$ and then the central charges $c^e_{R,L}$ in the electric picture. On the other hand, if we consider the perturbations carrying only magnetic charges, say $(d N_e,d N_m)=dN_m(0,1)$, the first laws help us to read the magnetic picture, in which the CFT is that of temperatures $T^m_{R,L}$ and central charges $c^m_{R,L}$. In general, we may consider the dyonic perturbations of the type $(d N_e,d N_m)=dN(a,b)$, with $a,b$ being two coprime integers. Then, from the first laws (\ref{e11}) we could get
\bea
&&\f{1}{2} d M=T_R d S_R + \O_R^N d N \nn\\
&&\phantom{\f{1}{2} d M}=T_L d S_L + \O_L^N d N,
\eea
with the intensive quantities conjugate to the integer-valued $N$ being $\O_{R,L}^N=a \O_{R,L}^e + b \O_{R,L}^m$. Using a similar method, we can easily get the dyonic picture, or $(a,b)$ picture, in which the CFT is that of the central charges
\be \label{e16}
c_{R,L}^{(a,b)}=a c_{R,L}^e+b c_{R,L}^m.
\ee

According to B\'ezout's lemma, for every pair of coprime integers $a,b$ there exist other pairs of coprime integers $c,d$ such that $ad-bc=1$. Thus, the $(a,b)$ picture CFT could be viewed as being generated from two elementary (1,0) and (0,1) pictures by a $SL(2,Z)$ transformation:
\be
\lt( \ba{c} c_{R,L}^{(a,b)} \\ c_{R,L}^{(c,d)} \ea \rt)
=\lt(\ba{cc} a & b \\ c & d \ea \rt)
\lt( \ba{c} c_{R,L}^{(1,0)} \\ c_{R,L}^{(0,1)} \ea \rt), ~~~
\lt(\ba{cc} a & b \\ c & d \ea \rt) \in SL(2,Z). \nn
\ee
In short, for the same dyonic RN black holes, there are many holographic descriptions, labeled by a pair of coprime integers $(a,b)$. These pictures are based on two elementary pictures and related to each other by $SL(2,Z)$ duality. Unlike the 4D Kerr-Newmann or 5D Myers-Perry black hole, this $SL(2,Z)$ could not be explained as the modular group of a torus. It has other interesting and profound origins.

\section{$SL(2,Z)$ As Electromagnetic Duality}

Using the electromagnetic duality, there is another way of generating different CFT dual pictures. For a given dyonic black hole, we could describe it in different gravity theories; i.e., different Einstein-Maxwell theories, which are related by the electromagnetic duality. We begin with an Einstein-Maxwell theory describing the black hole with the electric and magnetic charges given by (\ref{qeqm}), then we have the elementary (1,0) and (0,1) CFT pictures with the central charges given by (\ref{ce}), (\ref{cm}). Let us  change the Einstein-Maxwell theory by an $SL(2,Z)$ duality as described before:
\be \label{e2}
 \t'=\f{\a\t+\b}{\c\t+\d},  \hs{3ex}
\lt( \ba{c} N_{e'}\\N_{m'} \ea \rt)=
\lt(\ba{cc} \a & \b \\ \c & \d \ea \rt)
\lt( \ba{c} N_{e}\\ N_{m} \ea \rt).
\ee
Then we get a gravity theory whose (1,0)-picture central charges are
\be
c_{R,L}^{e'}=6 e' Q_{e'} Q^2.
\ee
Note that there is no prime in $Q^2$, because it is $SL(2,Z)$ invariant. From the transformation (\ref{e19}), we get the central charges
\be \label{e20}
c_{R,L}^{e'}=6Q^2 \lt[ \d e Q_e  - \c \lt( \f{Q_m}{e}-e Q_e \f{\th}{2\pi} \rt) \rt].
\ee
Comparing those results with (\ref{e16}), we find that we could get the same CFT picture  by setting $\d=a,\g=-b$.

We stress that (\ref{e16}) and (\ref{e20}) were obtained independently from two apparently different methods. The former was derived from the $SL(2,Z)$ duality of the CFT theories, while the latter was derived from the electromagnetic duality of the gravity theories. However, it turns out that the $SL(2,Z)$ duality in (\ref{e16})
originates from the electromagnetic duality in the gravity theory. Let us consider the dyonic  $(a,b)$-type perturbation with $(dN_e,dN_m)=dN(a,b)$ in the original theory. After the $SL(2,Z)$ transformations (\ref{e2}) with
\be
\lt(\ba{cc} \a & \b \\ \c & \d \ea \rt)
=
\lt(\ba{cc} d & -c \\ -b & a \ea \rt)
\ee
we have $(dN_{e'},dN_{m'})=dN(1,0)$, which is just a (1,0)-type perturbation in the ``new" gravity theory. Consequently, from the ``new" theory's point of view, we have a (1,0)-picture of the central charges from (\ref{e20}), in an exact match with  (\ref{e16}).

Furthermore, the electromagnetic duality in the gravity theory allows us to read the central charges by uplifting the solution to five dimensions. For a gauge potential with both electric and magnetic charges, it is not clear how to uplift the metric. However, using the duality, it is always possible to consider the configuration in a theory with only pure electric field. In this case, the metric and gauge potential may be uplifted to a 5D metric, from which the central charges could be computed.

\section{Conclusion and discussion}

In this paper, we investigated the holographic pictures of the four-dimensional RN black hole. It turned out that electromagnetic duality plays a subtle but important role in setting up generic holographic CFT duals of the black hole. It allowed us to study the same black hole background from different gravity theories, and find different pictures related to each other by the duality.

The notion of electromagnetic duality is new in the context of 4D Einstein-Maxwell theory, but is well known in string theory \cite{Cvetic:1995yp,Cvetic:1995yq}. The study in this paper indicates that the duality in the holographic descriptions of the black holes could be related to various dualities in string theory. It has been found in \cite{Horowitz:1996ay,Horowitz:1996ac} that there are classes of black holes whose entropies could be written as U-dual invariant forms. This suggests that these black holes should have multifold holographic duals \cite{Maldacena:1996ds}. It would be interesting to understand these duals more clearly from a thermodynamics points of view \cite{CZ}.

The thermodynamics of black holes has been well known for almost forty years, but it still brings us new surprise. It not only reflects the holographic nature of quantum gravity, but also encodes in itself the information of the holographic pictures and the symmetry among them. It certainly deserves more intense investigations from a microscopical point of view. We expect that string theory may shed light on this issue.

\vspace*{5mm}
\noindent {{\bf Acknowledgments}}
The work was in part supported by NSFC Grants No. 10975005 and No. 11275010. JJZ was also in part supported by the Scholarship Award for Excellent Doctoral Student granted by the Ministry of Education of China.
\vspace*{3mm}




\end{document}